\documentclass[letterpaper, 10 pt, conference]{ieeeconf}  

\IEEEoverridecommandlockouts                              

\usepackage{graphics} 
\usepackage{epsfig} 
\usepackage{mathptmx} 
\usepackage{times} 
\usepackage{amsmath} 
\usepackage{amssymb}  
\usepackage{url}

\title{\LARGE \bf
Head-up Displays (HUD) in driving}

\author{Enrique Ca\~{n}o$^{1}$, Pavel Gonz\'alez$^{2}$, Marcos Maroto$^{3}$ and Diego Villegas$^{4}$
\thanks{$^{1}$Enrique Ca\~no \ \  {\tt\small  100343797@alumnos.uc3m.es}}%
\thanks{$^{2}$Pavel Gonz\'alez {\tt\small 100325086@alumnos.uc3m.es}}%
\thanks{$^{3}$Marcos Maroto\ {\tt\small  marmarot@ing.uc3m.es}}%
\thanks{$^{4}$Diego Villegas \  {\tt\small  100336160@alumnos.uc3m.es}}%
}

\begin{document}

\maketitle
\thispagestyle{empty}
\pagestyle{empty}

\begin{abstract}
Head-Up Displays (HUDs) were designed \mbox{originally} to present at the usual viewpoints of the pilot the main sensor data during aircraft missions, because of placing instrument information in the forward field of view enhances pilots' ability
to utilize both instrument and environmental
information simultaneously. The first civilian motor vehicle had a monochrome HUD that was released in 1988 by General Motors as a technological improvement of Head-Down Display (HDD) interface, which is commonly used in automobile industry. The HUD reduces the number and duration of the driver’s sight deviations from the road, by projecting the required information directly into the driver’s line of vision. There are many studies about ways of presenting the information: \mbox{standard} one-earpiece presentation, \mbox{two-earpiece} three-dimensional audio presentation, visual only or \mbox{audio-visual} presentation. Results \cite{Durand} have shown that using a 3D auditory display the time of acquiring targets is approximately 2.2 seconds faster than using a one-earpiece way. Nevertheless, a disadvantage \cite{Gish95} is when the driver's attention unconsciously shifts away from the road and goes focused on processing the information presented by the HUD. By this reason, the time, the way and the channel are important to represent the information on a HUD. A solution is a context aware multimodal proactive recommended system \cite{Volvo} that features personalized content combined with the use of car sensors to determine when the information has to be presented.

\end{abstract}

\section{INTRODUCTION}
1.25 million people die approximately every year as a result of road traffic crashes \cite{who}. The HUD is a driver assistance system that reports timely the events that conceals a potential risk to both the driver and pedestrians  The first HUD was developed during World War I in order to help pilots to show down enemy planes, so it would no longer have to align the pilot’s head precisely with mechanical sights to make an accurate shot. These devices have expanded to include basic flight information like altitude, air speed, compass, and artificial horizon indicators.
\begin{figure}[h]
    \centering
    \includegraphics[scale = 0.3]{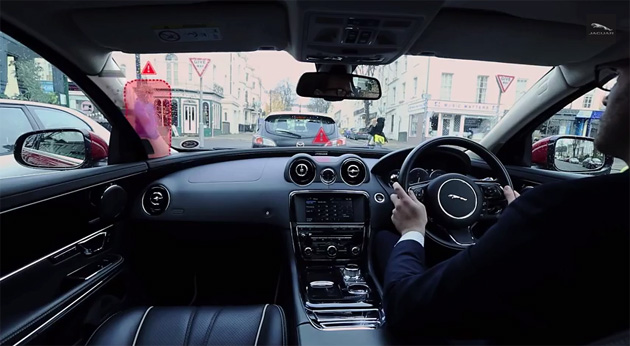}
    \caption{Jaguar Head-up display}
    \label{fig:HUD_Jaguar}
\end{figure}
In 1988, General Motors was the first company to install a monochrome HUD in a civilian motor vehicle. A car HUD system is formed by three main parts: (1) the combiner, the surface on which the image is projected; (2) the projector which generates the image and directs it toward the combiner and (3) the computing unit which processes data from different sources and handles the projection.
\\
Recent models of visual information processing suggest that visual attention can be focused on either Head-Up Displays (HUD) or on the world beyond them, but not on both simultaneously \cite{McCann93}. So it is important to have a suitable information presenting model to avoid distractions of driver’s attention and to manage the risk level of each indicator. This model has to be able to select the sensorial channel to use.
\begin{figure}[h]
    \centering
    \includegraphics[scale = 0.75]{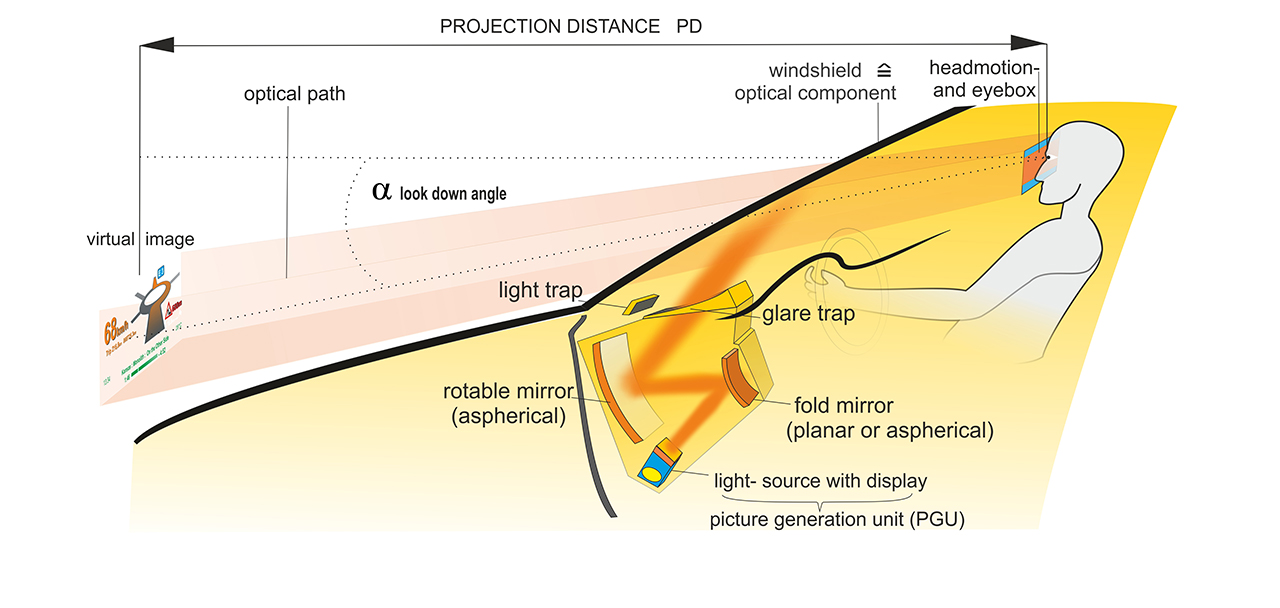}
    \caption{How a Head-up display works}
    \label{fig:HUD_Continental}
\end{figure}
On this paper, we discuss about the applications of a HUD on a civilian motor vehicle and the different advantages that can be balanced with disadvantages on a real driving situation. Finally, we are talking about some commercials products, manufacturers of HUD and several future development and experimental designs using augmented reality on HUD.
\section{APPLICATIONS}
Most applications where Head Up Displays \cite{Wood2006} are used, are related to professional issues, mainly in military aircraft, where they were used for the first time. However, nowadays, it is increasing the number of fields in which HUDs are being used like automobile, augmented reality in video games, etc. As Intelligent Transport Systems case of study, we are going to focus on automobile applications, taking into account new and experimental applications of HUD and giving a brief introduction about HUDs installed on aircraft as a first step.
\subsection{Aircraft}
First HUDs used in aircraft \cite{DigitalAvionics} appeared as an aid mechanism for military pilots during missions. These systems represent powerful information to the pilot in order to allow them to see this information without removing the view from the front.\\
Some of the basic information that it is represented in the frontal panel of the aircraft is airspeed, altitude, the horizon line, heading, turn/bank and slip/skid indicators. In addition, some HUD mounted on aircraft can include some extra information like bore-sight or waterline symbol, flight path vector (FPV) or velocity vector, acceleration, etc. The main applications of HUDs in aircraft's field are:
\begin{itemize}
    \item Military aircraft applications: apart from the information represented in basic HUDs, there are some others specific applications related to sensor data and weapon systems.
    \begin{itemize}
        \item Target designator (TD) indicator: trough the HUD and with the data acquired from the inertial and radar sensors, it is represented a cue around an aerial or ground target allowing the pilot to identify it.
        \item Weapon seeker or sensor line of sight: represents the direction to which a weapon or a seeker sensor is pointing.
        \item Weapon status: indicates the selected weapon, available weapon, arming and everything related to the state of the aircraft's weapons.
        \item Closing velocity with target (Vc).
        \item Range to target, waypoint, etc. It shows the distance or range to important places or points.
    \end{itemize}
    \item Civil aircraft applications \cite{Goteman}: the most important application on civil aircraft is related to take-offs and landings with reduced visibility, as an assistant to the pilot.
    \item Enhanced vision systems (EVS) and Synthetic vision systems (SVS) \cite{Arthur2005}: EVS systems are included in advanced systems that incorporate information from aircraft based sensors (e.g., near-infrared cameras, millimeter wave radar) to provide vision in limited visibility environments. SVS systems use high precision navigation, attitude, altitude and terrain databases to create realistic and intuitive views of the outside world. EVS and SVS are commonly used as a very important complement part of HUD in recent military aircraft.
    \begin{figure}[h]
    \centering
    \includegraphics[scale = 1.5]{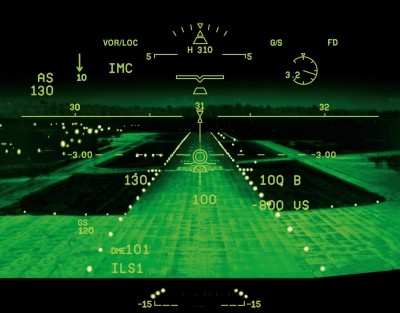}
    \caption{EVS infrared image with HUD information.}
    \label{fig:HUD_EVS}
    \end{figure}
\end{itemize}
\subsection{Automobile}
On cars, some companies like BMW, Audi, Mercedes Benz or Hyundai are betting for HUD technology. Although there are some similarities between the design parameters of HUDs installed on aircraft and HUDs installed on automobiles, automotive HUDs differs from aircraft HUDs because the information represented on automobile HUDs do not need to provide the driver information about the location with respect to the road, as HUDs installed on aircraft give to the pilot, though, because the driving environment usually gives enough information to the user.  \\The information represented on the windshield of cars is mostly related to:
\begin{itemize}
    \item Characteristics of the road like speed limit of the highway.
    \item Navigation (e.g. maps displays). 
    \item Automobile's operations  like the speed of the car, distance to a goal or information about the path to a goal(GPS) and turns of the car. 
\end{itemize}
The current components of an automotive HUD can vary in several ways, but all HUDs installed on automobiles have the same main elements: an image source, a set of lenses and mirrors to reflect, refract, focus and magnify the HUD image, and a combiner surface (where the image is projected, typically over the windshield).
\\
The main applications of automobile HUDs \cite{Harrison1994} are related to comfortableness and safety to the driver. The main ones are:
\begin{itemize}
    \item Supply information to the driver about the main internal information of the vehicle (e.g. remaining fuel, adverts about the tires, etc.).
    \item Information about non-expected events like accidents, state of the roads or bad weather conditions.
    \item Information about the navigation of the vehicle like its position in a map or the way to go during a trip.
    \item Information related to bad or unwanted behaviours of the driver during driving like sharp turns or drowsiness warning.
\end{itemize}
On the other hand, several companies like Garmin or Pioneer \cite{JayAlabaster}, have developed new commercial displays that can be installed as external HUD devices and show some information about the car's navigation system (GPS) on the windshield of the car. The difference between the HUD installed internally in new cars and HUD installed on an external way is that internal HUD can use every information of the car (navigation, internal sensors, cameras in new cars) while external devices like the ones developed by Garmin or Pioneer only use navigation information because it hasn't got any way to access to the internal information of the car. \\
In addition, new cars can include night vision information \cite{Groves}, equipping cars with an infrared camera for viewing roadway conditions in terms of a thermal image, and outputs a video signal to a head up display (HUD) which projects the camera view to the operator via the windshield or other combiner to display a virtual image in the operator's field of view. The HUD is configured to magnify the image to the same size as the visual or real scene, and compensates for camera and windshield distortion. The virtual image is presented above or below the real scene or may be superimposed on the real scene. A video processor allows selection of only the warmest objects for display.
\begin{figure}[h]
    \centering
    \includegraphics[scale = 0.15]{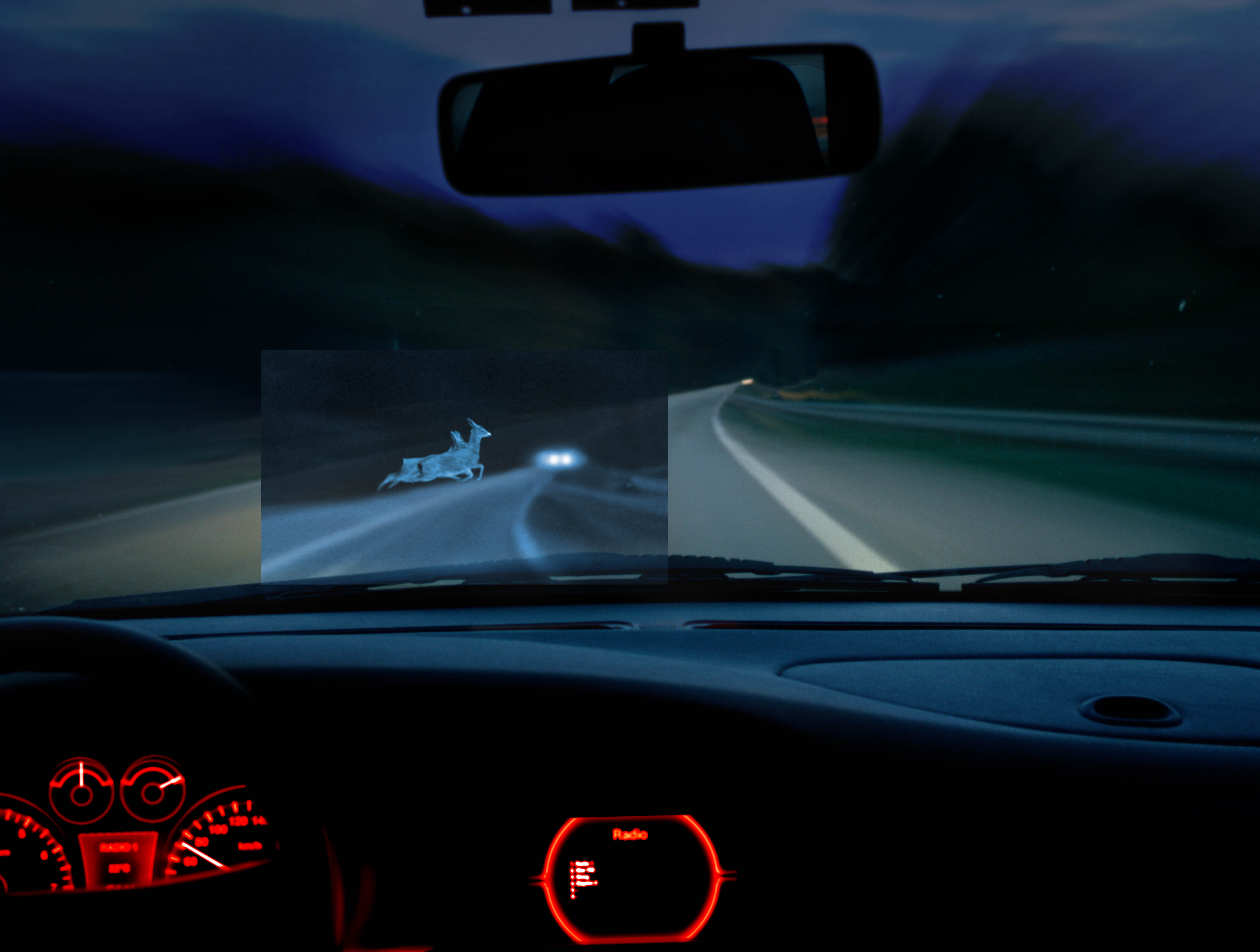}
    \caption{Night vision infrared camera information as HUD}
    \label{fig:HUD_Night}
    \end{figure}
HUDs have been installed successfully on aircraft with a cost of \$100.000, while in automobiles, reducing some of their characteristics related to the range of view and the image source, basic HUDs installed on cars can have its cost reduced up to \$100 per unit.\\
As an experimental development, HUDs have been also placed in rear-view mirrors of the car to provide drivers of extra information of the rear view like the distances while the car is being overtaked by another vehicle or when it is the own car provided by this HUD system which overtakes another car. Also, some companies are integrating HUDs on helmets \cite{Nicholson2011} for motorcyclist that represent useful information like black spots in roads in order to reduce the number of accidents, advertising messages of speed, etc.
\begin{figure}[h]
    \centering
    \includegraphics[scale = 0.35]{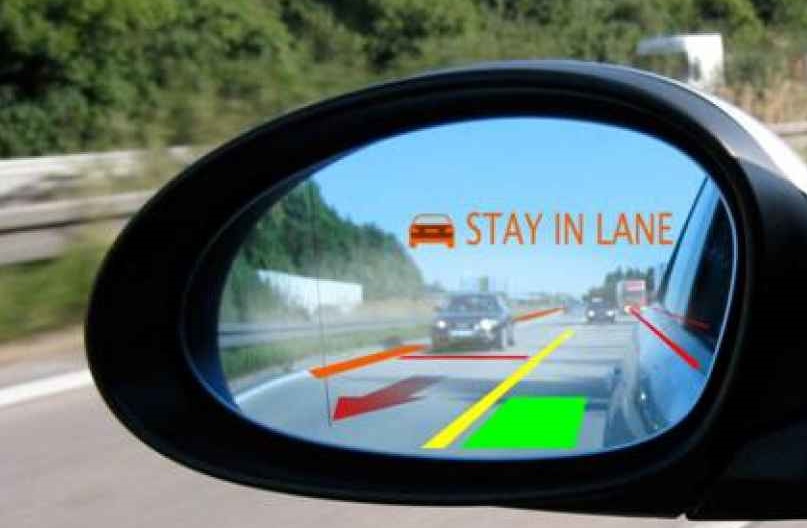}
    \caption{Night vision infrared camera information as HUD}
    \label{fig:HUD_rear-view}
\end{figure}

\section{ADVANTAGES AND PROBLEMS}
A head-up display (HUD) is a revolutionary ADAS (Advanced driver assistance system) that can change the way we drive in a smart, safe and technological way. \cite{url_howtogeek}
\subsection{Advantages}
The use of a HUD in a car origin several advantages to take into account:
\begin{itemize}
\item	Head up display will replace the indicator cluster in dashboard on common car, also known as Head-Down Display (HDD). The data that will be shown on HUD is: engine speed, vehicle speed, engine temperature, fuel capacity, pedestrian and vehicle warning, object avoidance guidance.\cite{Wuryandari13}
\begin{figure}[h]
    \centering
    \includegraphics[scale = 0.165]{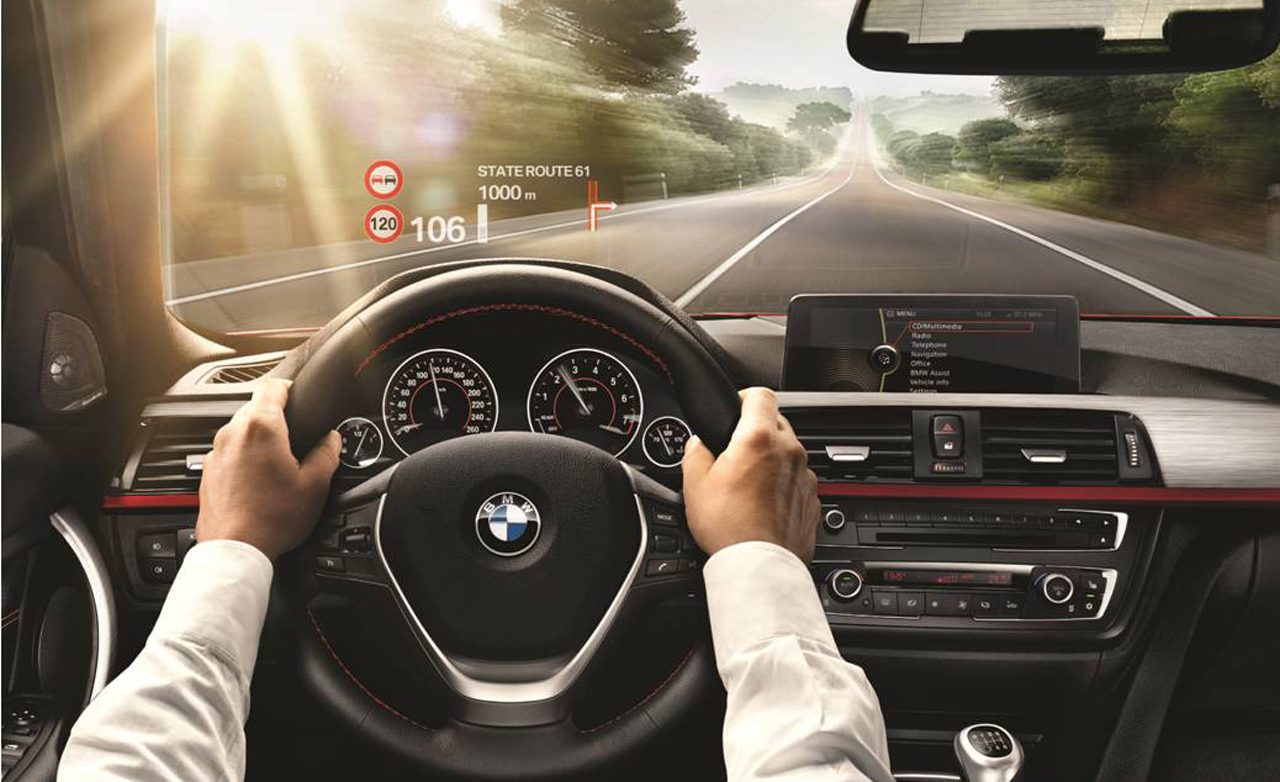}
    \caption{BMW Head-up display}
    \label{fig:HUD_BMW}
\end{figure}
\item	A HUD is a device that can help to avoid several factors that cause accidents on the road. Those factors are human error, vehicle error, and the accident itself which cannot be avoided. The sources of accident because of human error are sleepy, drunk, or using mobile phone while driving. A HUD is the necessary device to solve those problems.
\item To upgrade the efficiency and safety, a HUD system must provide drivers with large amounts of information from many categories (e.g. route guidance/navigation, traffic signs, cargo/road/vehicle conditions), and select the best way to display this information; important considerations include having a user-friendly system, since a driver’s capacity to process this information is a key factor in its acceptance and use.\cite{Liu04}
\item	In comparison with HDD( Head-Down Display):
    \begin{itemize}
    \item	In terms of response time to an urgent event, it was faster with the HUD (with a low driving load—head-up vs. head-down: 1.0073 vs. 1.8684 s; with a high driving load—head-up vs. head-down: 1.3235 vs. 2.3274 s) and speed control was more consistent (having low speed variations) than with the HDD. \cite{Liu04}
    \item	In addition, using the HUD caused less mental stress for the drivers than the HDD.\cite{Collins99}
    \item	As head-down displays are integrated into the vehicle's control dashboard, by using HUD the frequency and duration of glances towards the display are reduced by presenting information directly on the windshield in the driver's field of vision. \cite{Ablassmeier07}
    \item In comparison to the HDD, most HUD research has focused on providing information on speed limit restrictions, the driver’s reactions to accidents and psychological condition. It has also been found that automobile speed is maintained at a more consistent level (120 km/h within speed limits) \cite{Sojourner90}, that drivers were more aware of the speed of their vehicles and more closely adhered to the posted speed limit, while using an HUD. In addition, the speed of drivers having an HUD is, on average, faster than that of drivers having to look head-down at the dashboard.
    \item The head-up display (HUD) reduces the number and duration of the driver’s sight deviations from the road, by projecting the required information directly into the driver’s line of vision. This allows drivers to receive information without lowering their gaze, thus avoiding attention gaps that result from them taking their eyes off the road to look down at the information on a HDD \cite{Kaptein94}. In this way, the driver can easily keep his driving under control and can quickly respond to information relating to the road environment from the in-vehicle communication system.
    \end{itemize}
\item In addition, HUD technology has seen many improvements. For example, HUD's are now capable of adjusting to ambient light, so the images are just as clear whether you are driving at day or night.\cite{Hawaii}
\item HUDs allow taking advantage of advances in technology in the communication process between the driver and the car. Nowadays, vehicles have fuel-efficient hybrid engines, proximately sensors, windshield wipers that can detect rain, built-in multimedia entertainment, and all-wheel drive systems that adjust power in real-time, but the interaction between the driver and the car has not changed significantly.\cite{Ingman05}
\item A HUD could also be used to: Mobile phone access (phone call, SMS, multimedia); display navigation information, traffic information, and mobile phone interaction; handle the voice command for mobile phone interaction and passenger condition monitoring.\cite{Jakus15}
\item	Some authors pointed out that if a driver’s gaze leaves the road for longer than 2s then traffic accident risk is significantly increased. This ‘attention-away-from-the road’ situation is one of the main factors causing danger on the roads.\cite{Zwhalen88}
\end{itemize}

\subsection{Problems}
   \begin{itemize}
\item	In comparison with HDD( Head-Down Display) \cite{Liu04}:
    \begin{itemize}
        \item	HUD was less comfortable for first-time users to become familiar with; with a high driving load, however, the difference between the two displays was not significant. 
        \item	If a HUD is used to show different information apart from driving one, as for example: changing the music playlist, sending a restaurant reservation or checking the email while you’re driving, the distraction increased in an exponential form. A HUD must show information related with the car and its environment. A HUD has to guarantee the safety.\cite{Ward94}
        \item	Several authors affirmed that drivers felt that the HUD required a higher mental effort and produced higher mental demand workload ratings
    \end{itemize}
\item However, cognitively switching between two sources of information, i.e., the HUD and traffic, still poses a problem, especially in high-workload situations. The so-called cognitive or attention capture, i.e., when the driver's attention subconsciously shifts away from the road and becomes focused on processing the information presented by the HUD, has been identified as one of the disadvantages of HUDs \cite{Prinzel04}. The resulting perceptual tunneling may lead to a delayed reaction or a complete absence of response to situational changes in the environment. Researchers have proposed to address this issue by utilizing auditory interfaces \cite{Sodnik08}. They report on the greater attention capturing properties of the auditory channel and the desire to keep attention focused longer on a complex auditory task to prevent a loss of information from the driver's working memory.
\end{itemize}  

\section{AVAILABLE COMMERCIAL APPLICATIONS}

Commercial use of the HUD is already taking off since early 2000's, but in the next years is going to be more common to see these systems in different classification of uses. This chapter will describe the three main classifications of the available systems, the one that comes with the vehicle from the manufacturers, the smartphone app whose screen is reflected on the windshield and the third-party devices that can be installed on any car.  

\subsection{Already installed}
The most widespread use of HUDs is being deployed in top of line car models, like BMW, Mercedes, and Audi. Nevertheless, more manufacturers include this as an option for safety or for comfort every year. These are HUDs that are already installed as a part of the car, which means they can pull data from everything that’s actually happening from inside the engine in real-time, without the need for any GPS-assisted guesswork.
\begin{itemize}
    \item \textbf{BMW }projects relevant driving information directly into the driver’s line of sight. Head-Up Display can be recognized by a small square depression on the dashboard. This contains a projector and a system of mirrors that beams an easy-to-read, high-contrast image onto a translucent film on the windscreen, directly in the line of sight. The image is projected in such a way that it appears to be about two meters away, above the tip of the bonnet, making it particularly comfortable to read. The full-color BMW Head-Up Display in many newer BMW models makes the car even more comfortable for the driver. More colors mean it’s easier to differentiate between general driving information like speed limits and navigation directions and urgent warning signals. Important information like ‘Pedestrian in the road’ (in conjunction with the optional BMW Night Vision) is now even more clear and recognizable. In addition to the speed, which is displayed permanently, the BMW Head-Up Display can also show other content depending on model and equipment \cite{BMW}. HUD can be found starting in the BMW Series 3.
    \item \textbf{Mini Cooper}: The screen is installed on the dashboard and is positioned at the height of the steering wheel. All of the information is therefore in the driver's direct line of vision. The Head-Up Screen has an integrated smartphone interface. As a result, information from smartphone apps can also be displayed. Garmin, as a navigation specialist, offers a matching navigation application. As a result, next to the relevant vehicle data, the Head-Up Screen also displays navigation information. All of the settings can also be carried out via the free-of-charge Head-Up Screen app. During the hours of darkness, the display adjusts the brightness automatically. \cite{Mini}. It can be adding to almost the all Mini series.
    \item \textbf{Chevrolet:} HUDs already available on the Chevrolet Corvette, and since the 2011 Camaro, it includes: Vehicle speed, Tachometer, Compass, Outside air temperature Manual Paddle Shift Gear Indicator (if equipped), OnStar Turn-by-Turn, Audio functions, Phone information, Selected gear, Turn signal indicators, High-beam indicator signal, Vehicle messages.
    \item \textbf{Peugeot 3008:} The French manufacturer Peugeot has chosen a slightly different HUD system than other brands, instead of projecting directly onto the windshield, Peugeot does on a small panel of translucent polycarbonate behind the dashboard, which can be hidden if desired \cite{Autofacil}. System allows to adjust the brightness of the projected image, either manually or automatically (with light sensors) to the lighting conditions outside, to see the text clearly whatever the circumstances. HUD Peugeot allows to set the safe distance from the vehicle in front of the car, so the indicator on the HUD flashes because there is "area danger”.
\end{itemize}
Other manufacturers and the model with HUD included are: Toyota Prius, 	Mazda 3,	Jaguar XE,	Volvo XC90,	Lexus RX,	Hyundai Genesis,	Rolls Royce,Mercedes-Benz Class C and Class S Sedan and Coupe,Audi and	General Motors SUVs, among others.

\subsection{Smartphones apps}
On the other hand, there are software developments that emulated HUD for smartphones users that want to try this kind of technology. 
\\
Lately there have been a collection of apps popping up both in the iTunes and Google Play app stores which promise all the same functionality you’d expect from a traditional HUD setup, except without any of the hassle of installation or getting all the wiring right between incompatible cars. The apps work by brightly displaying a reverse image on the windshield when the phone is mounted on top of the dashboard itself. Some of then are:
\begin{itemize}
    \item \textbf{ Navier HUD} is an application available on Android phones. Navier HUD uses the HUD (Head-up display) concept to help drivers without having to take their eyes off the road. The driving information is projected on to the windshield when using Navier HUD while the phone lie down in front of the windshield. Navier HUD has a simple navigation function to hint people toward a selected destination just follow the routing instruction. For the use of HUD, driving information and instruction are designed as simple symbols to help drivers easy to recognize. \cite{PlayStore}
    \item \textbf{HUDWAY }is a unique application for smartphone users that allows drivers to focus on the road, keep their hands on the wheel and mind on driving. Now drivers don't have to interpret audio commands or complicated maps and symbols.
    Current navigation systems are inherently unsafe by distracting a driver’s attention away from the road. This is especially true when driving in low visibility conditions — rain, fog, heavy snow or darkness.\cite{PlayStore}
    \item \textbf{Sygic}: See all navigation instructions projected directly into your line of sight - the windshield. This allows you to process instructions faster and keep your attention on the road. Head-up Display works without any additional accessories or devices. It projects all important information such as turn indicator arrows, distance to the next turn, your current speed and speed limit. Warnings for speed cameras and estimated time of arrival are also included.\cite{PlayStore}
    \begin{figure}[h]
    \centering
    \includegraphics[scale = 0.22]{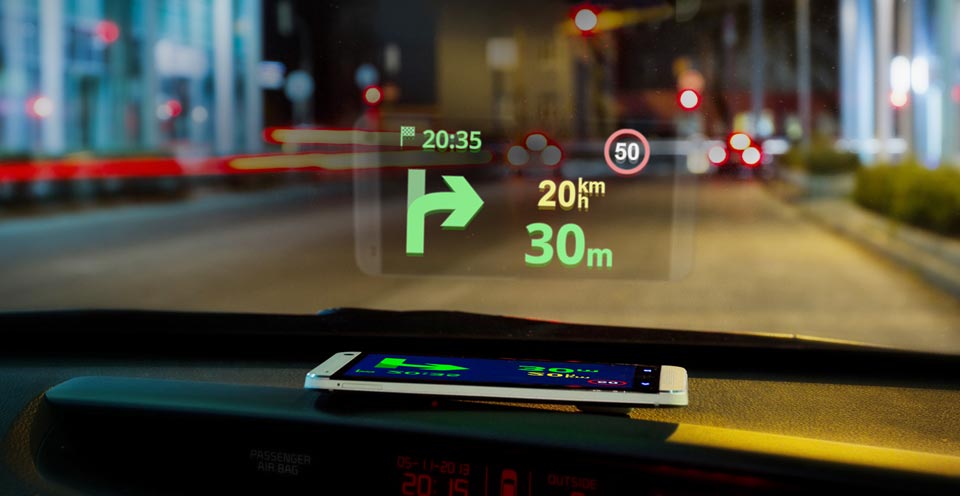}
    \caption{Sygic Head-up display}
    \label{fig:HUD_Sygic}
\end{figure}
\end{itemize}
All this apps can be tested but to fully use them, is about 9\$ purchase.
\subsection{Third-party device}
Most third-party HUDs work by linking to either your phone’s internal GPS or finding a signal of their own from a satellite to guesstimate how fast your car is going at any given time, and display the information back on the windshield \cite{Viso}.For now, HUDs made by individual companies for aftermarket use are only capable of displaying a rudimentary speedometer, often in mono-chromatic colors which aren’t all too pleasing for the eye to be staring at for hours on end. Below, there are some manufacturers:
\\
\\
\textbf{Garmin HUD+ }beams navigation information onto your windshield for easy viewing while driving. It receives data wirelessly from your compatible smartphone running the free Garmin HUD app. 
The device offers far more navigation detail than other portable HUD systems, displaying turn indicators, distance to your next turn, current speed and speed limit, and estimated time of arrival. HUD+ lets you know what lane to be in for your next 
maneuver as well as alerts you if you exceed the speed limit.\cite{Garmin} It costs €150.
\\
\textbf{Other Manufacturers}
\\
In the market we can find several manufacturers that offer self-mounting devices HUD like: 
\begin{itemize}
    \item Kshioe 5.5" Q7 Universal GPS HUD
  \item Eoncore New Universal 5.5" Car A8 Hud Head Up Display with OBD2 Interface
  \item Generic X6 3" Universal Multi-function compatible with OBD2 System Model Cars
\end{itemize}

All of them start at €50, and take 1 hour of installation.
\subsection{Future commercial developments}
\subsubsection{Wind shield display}
Autoglass® 2020 vision: the future of the car windscreen
Augmented reality - intelligent car windscreens that provides real-time visual information to drivers are only a few years away, according to Autoglass®.
\\
The UK's leading vehicle glass repair and replacement specialist is predicting that the car windscreen will soon function as an interactive visual display for everything going on in and around the car.
\\
To demonstrate how the windscreen of the future would work, Autoglass® has produced a video of how driving around London in 2020 might look.
\\
By 2020, the windscreen will display key information about speed, fuel and any issues with parts of the car. It will also combine visual sensors with augmented reality, online maps and GPS technology to provide drivers with live, visual information about the places and hazards around them.\cite{Autoglass}
\subsubsection{Holograms systems}
Aida 2.0 project aims at estimating a driver's likely destination based on collective mobility patterns in the city and individual profile information -such as past riding behaviours and online calendar entries-. Aida proposes an unprecedented interface that allows the driver access this information. In an effort to bring the virtual augmented map closer to the actual physical city sean through the windshield, Aida's interface exploits most unused area on the dashboard to establish a direct connection between the actual street and ist representation on the digital map.\cite{Aida}
\subsubsection{Helmet}
In order to provide further optimized this technology in the future , BMW is working on a helmet with an innovative feature Head- Up Display . This allows the projection data directly in the driver's field of vision so that you no longer have to consult the instrument panel and can fully concentrate on traffic.
\\
All indicators freely programmable display to give the best possible support for pilot safety. The idea is that only the information that is useful and relevant to the situation of each particular time appears.
\\
It includes augmented reality so drivers don't have to look down at their gauges. It's like the Google Glasses 2.0 for motorcycle helmets.
\\
The technology for motorcyclists involves a color display about as thin as eyeglasses that fits within helmets. It magnetically clips into the helmet in front of the riders' eye, avoiding any disruption to their line of sight while accommodating eyeglasses.
\\
BMW is the first company to license the DigiLens technology. The automaker has previously used the technology in its MINI augmented vision glasses, which debuted at the Auto Shanghai show in China on April. So exploring the technology for its motorcycles is a natural extension, according to Richter.\cite{Helmet}

\section{CONCLUSIONS}
HUD devices are Advanced Driver Assistance Systems (ADAS) whose objective is to enhance driving safety and efficiency thank to the information obtained from the environment and the vehicle.
\\
HUD provides the driver a big amount of information avoiding losing the attention from the road because all the information is shown on the windshield.
\\
Its use has grown in an exponential way and in the next years, it is expected to be installed on every car.
\\
Nowadays, there are too many automotive companies that have developed commercial HUD systems like BMW, Jaguar, Mercerdes or Peugeot among others. 

\addtolength{\textheight}{-10cm}   




\end{document}